\documentstyle[12pt]{article}
\begin{document}
\begin{center}
POINCARE ANOMALY IN PLANAR FIELD THEORY\\
\vskip 5cm
Subir =
Ghosh\footnote{email:sghosh@isical.ac.in;subir@boson.bose.res.in}\\
\vskip 2cm
P. A. M. U., Indian Statistical Institute,\\
203 B. T. Road, Calcutta 700035, India.
\end{center}
\vskip 4cm
Abstract:\\
We show the presence of Poincare anomaly in Maxwell-Chern-Simons theory =
with an explicit mass term, in 2+1-dimensions.
\newpage
The ubiquitous use of 2+1-dimensional field theories in condensed matter =
systems, where the dynamics normal to a plane is severely restricted, =
has enlarged the scope of lower dimensional physics from being just a =
toy model of the 3+1-dimensional world.
The topologically massive Maxwell-Chern-Simons (MCS) gauge theory
 was first thoroughly analysed by Deser, Jackiw and Templeton (DJT) in =
their seminal work \cite{djt}, where the subtle interplay between =
Poincare invariance and an unambiguous determination of the spin of the =
excitations in a vector theory was revealed. It was shown that correct =
space-time transformation of the gauge invariant observables, such as =
electric and magnetic fields, were induced by Poincare generators which =
obey an anomalous algebra among themselves. However, a phase =
redefinition of the creation and annihilation operators removed the =
commutator anomaly and yielded the spin contribution in a single stroke. =
In the present Letter, we consider the MCS model with an explicit mass =
term, {\it i.e.} the MCS-Proca (MCSP) model. We show that in the =
presence of two mass scales, the topological one $(\mu )$, generated by =
the Chern-Simons term, and the explicit (gauge symmetry breaking one) =
one $(m)$, the anomaly in the Poincare transformationsin the =
eletromagnetic fields can not be removed, even though the equations of =
motion are manifestly Lorentz covariant. This is our main result.

The MCSP model was studied previously in \cite{pk}. It also appears =
naturally in the large fermion mass limit of the bosonization of gauged =
massive Thirring model \cite{g}. In \cite{pk}, it was argued that the =
self-dual factorisation of the equation of motion leads to two self and =
anti-self dual excitations of different masses, thereby accounting for =
the parity violation induced by the topological term. Recent results =
\cite{bk} indicate, in the path integral formalism, that the above naive =
conclusions are invalid. As shown in \cite{bk}, the fact that MCSP model =
is a result of a fusion between self and anti-self dual models explains =
the self dual factorisation of the equation of motion. But in the =
process, the self and anti-self dual property of the MCSP model is no =
longer manifest.
This controversy demands an indepth analysis of the model.

The MCSP model, with the metric being $g_{\mu\nu }=3Ddiag(+--),~\epsilon =
_{12}=3D1$, is
\begin{equation}
{\cal L}_{MCSP}=3D-{1\over 4}A_{\mu\nu }A^{\mu\nu }
+{{\mu }\over 4}\epsilon_{\mu\nu\lambda }A^{\mu\nu }A^\lambda
+{{m^2}\over 2}A_\mu A^\mu ,~~A_{\mu \nu }=3D\partial _\mu A_\nu =
-\partial _\nu A_\mu .
\label{j}
\end{equation}
Taking $m^2=3D0$ reproduces the MCS theory, which being a gauge theory =
is amenable to gauge fixing conditions. This simplifies the model =
considerably and makes the field content transparant. We also try to =
implement similar parametrizations as in \cite{djt} and hence convert =
the above gauge non-invariant theory to a gauge invariant one by the =
Stuckelberg prescription,

\begin{equation}
{\cal L}_{St}=3D-{1\over 4}A_{\mu\nu }A^{\mu\nu }+{{\mu }\over =
4}\epsilon_{\mu\nu\lambda }A^{\mu\nu }A^\lambda +{{m^2}\over 2}(A_\mu =
-\partial _\mu \theta )(A^\mu -\partial ^\mu \theta ),
\label{k}
\end{equation}
where $\theta $ is the Stuckelberg field. We define the conjugate =
momenta \cite{djt} and the Poisson bracket algebra as,
$$
{{\partial {\cal L}_{St}}\over {\partial \dot A^i}}\equiv \Pi ^i=3D-\dot =
A_i+\partial _iA_0-{\mu \over 2}\epsilon_{ij}A_j;
~~{{\partial {\cal L}_{St}}\over {\partial \dot A^0}}\equiv \Pi =
^0=3Dm^2\theta ;~~{{\partial {\cal L}_{St}}\over {\partial \dot \theta =
}}\equiv \Pi _\theta =3Dm^2\dot \theta ,$$
\begin{equation}
\{A_\mu (x),\Pi _\nu (y)\}=3D-g _{\mu \nu }\delta (x-y),~ \{\theta =
(x),\Pi _\theta (y)\}=3D\delta (x-y) .
\label{l}
\end{equation}
The Hamiltonian is
$$
{\cal H}_{St}=3D\Pi ^\mu \dot A^\mu + \Pi _\theta \dot \theta -{\cal =
L}_{St} $$
$$
=3D{1\over 2}\Pi _i^2+{1\over 4}A_{ij}A_{ij}+({{m^2}\over 2}+{{\mu =
^2}\over 8})A_i A_i=20
-{\mu \over 2}\epsilon_{ij}\Pi _i A_j $$
\begin{equation}
+{1\over {2m^2}}\Pi _\theta ^2
 +{{m^2}\over 2}\partial _i\theta \partial _i\theta +m^2(\partial =
_iA_i)\theta
-A_0(\partial _i\Pi _i+{\mu \over 2}\epsilon _{ij}\partial =
_iA_j+{{m^2}\over 2}A_0),
\label{m}
\end{equation}
where a total derivative term has been dropped. The two involuting first =
class constraints, (in the Dirac sense of classification), are
\begin{equation}
\chi _1\equiv \Pi _0-m^2\theta,~~\chi _2\equiv \partial _i\Pi _i+{\mu =
\over 2}\epsilon _{ij}\partial _iA_j+m^2 A_0 +\Pi _\theta .
\label{n}
\end{equation}
 The unitary gauge,
$\phi _1\equiv \Pi _\theta;~~\phi _2\equiv \theta $,
establishes gauge equivalence between the embedded model and the =
original MCSP model. This ensures that in the gauge invariant sector, =
results obtained in any convenient gauge will be true for the MCSP =
theory.
We invoke the rotationally symmetric Coulomb gauge \cite{djt}
\begin{equation}
\psi _1\equiv A_0;~~~\psi _2\equiv \partial _i A_i .
\label{o}
\end{equation}
The $(\chi _i,~\psi _j )$ system of four constraints are now second =
class, meaning that the constraint algebra metrix is invertible. The =
Dirac brackets, defined in the conventional way are given below,

$$
\{A_i(x),\Pi _j(y)\}^*=3D(\delta _{ij}-{{\partial _i\partial _j}\over =
{\nabla ^2}})\delta(x-y);~~\{\Pi_i(x),\Pi_j(y)\}^*=3D-{\mu \over =
2}\epsilon _{ij}\delta (x-y) $$
\begin{equation}
\{\Pi _i(x),\theta (y)\}^*=3D{{\partial _i}\over {\nabla =
^2}}\delta(x-y);~~\{\Pi _i(x),\Pi _0 (y)\}^*=3D-m^2{{\partial _i}\over =
{\nabla ^2}}\delta(x-y).
\label{p}
\end{equation}
The remaining brackets are same as the Poisson brackets. The reduced =
Hamiltonian in Coulomb gauge is
\begin{equation}
{\cal H}_S=3D{1\over 2}\Pi _i^2+{1\over 2}\partial _iA_j\partial =
_iA_j+({{m^2}\over 2}+{{\mu ^2}\over 8})A_i A_i=20
-{\mu \over 2}\epsilon_{ij}\Pi _i A_j+
{1\over {2m^2}}\Pi _\theta ^2
 +{{m^2}\over 2}\partial _i\theta \partial _i\theta .
\label{q}
\end{equation}
Although somewhat tedious, it is straightforward to verify that the =
following combinations,
$\phi =3D((\epsilon _{ij}\partial _i A_j),(\epsilon _{ij}\partial _i\Pi =
_j),\Pi _\theta , \theta )$ obey the higher derivative equation
\begin{equation}
(\Box +M_1^2 )(\Box +M_2^2)\phi =3D0;~~M_1^2(M_2^2)=3D{1\over =
2}[2m^2+\mu ^2\pm \mu {\sqrt {\mu ^2+4m^2}}].
\label{u}
\end{equation}
The spectra agrees with \cite{pk}. Note that
for $\mu ^2 =3D0$, the roots collapse to $M_1^2=3DM_2^2=3Dm^2$, which is =
just the Maxwell-Proca model, whereas for
$m^2=3D0 $ the roots are $M_1^2=3D\mu ^2,~M_2^2=3D0 $, indicating the =
presence of only the topologically massive mode, since the Stuckelberg =
field $\theta $ is no longer present.

Prior to fixing the $\psi _2 $ gauge, the gauge invariant sector is =
identified as,
\begin{equation}
E_i=3D-\Pi _i+{\mu \over 2}\epsilon _{ij}A_j;~~ B=3D-\epsilon =
_{ij}\partial _iA_j;~~\Pi _\theta ;~~A_i+\partial _i\theta ,
\label{y}
\end{equation}
where $E_i$ and $B$ are the conventional electric and magnetic field.
In the reduced space, the Hamiltonian and spatial translation generators =
are gauge invariant,
$${\cal H}_{St}=3D{1\over 2}(E_i^2+B^2+{{\Pi _\theta ^2}\over =
{m^2}}+m^2(A_i+\partial _i\theta )^2),$$
\begin{equation}
{\cal P}_{St}^i=3D-\epsilon _{ij}E_jB-\Pi _\theta (A_i+\partial _i\theta =
).
\label{x}
\end{equation}
Defining the boost transformation as $M^{i0}=3D-t\int d^2x{\cal =
P}_{St}^i(x)+\int d^2xx^i{\cal H}_{St}(x)$, the Dirac brackets with the =
gauge invariant variables are easily computed. They will contain non =
canonical pieces in order to be consistent with the constraints. =
However, changing to a new set of variables by the following canonical =
transformations,=20
\begin{equation}
Q_1(Q_2)=3D{1\over {{\sqrt {-2\nabla ^2}}}}[\epsilon _{ij}\partial =
_iA_j\pm {1\over m}\Pi _\theta ];~~
P_1(P_2)=3D[{1\over {{\sqrt {-2\nabla ^2}}}}\epsilon _{ij}\partial _i\Pi =
_j\mp
{m\over 2}{\sqrt {-2\nabla ^2}}\theta ],
\label{w}
\end{equation}
we can convert our system to a nearly decoupled one.
Passing on to the quantum theory, the redefined variables satisfy the =
canonical algebra,
\begin{equation}
i\{P_i,Q_j\}=3D\delta _{ij}\delta (x-y);~~\{Q_i,Q_j\}=3D\{P_i,P_j\}=3D0.
\label{xy}
\end{equation}
The electric and magnetic fields and the translation generators are =
rewritten as,
\begin{equation}
B=3D-{{\sqrt {-2\nabla ^2}}\over 2}(Q_1+Q_2);~~
E_i=3D-{1\over {\sqrt {-2\nabla ^2}}}[\epsilon _{ij}\partial =
_j(P_1+P_2)+(\mu + m)\partial _iQ_1+(\mu -m)\partial _iQ_2],
\label{z}
\end{equation}
$$
H_{St}=3D\int d^2x[{1\over 2}(P_1^2+\partial _iQ_1\partial =
_iQ_1+M_1^2Q_1^2)+{1\over 2}(P_2^2+\partial _iQ_2\partial =
_iQ_2+M_2^2Q_2^2)+{{\mu ^2}\over 2}Q_1Q_2]$$
\begin{equation}
P^i_{St}=3D\int d^2x[P_1\partial ^iQ_1+P_2\partial ^iQ_2]
\label{ff}
\end{equation}

In order to drive home the peculiarities of MCSP theory,
let us briefly consider the special cases, $m^2=3D0$ or $\mu ^2=3D0$.
In the former limit, giving the MCS theory, as we noted before, $\theta =
$ field is absent, which makes the $(Q_1,~P_1)$ pair identical to the =
$(Q_2,~P_2)$ pair, leading to the following relations, with =
$i[p(x),q(y)]=3D\delta (x-y)$,
$$B=3D{\sqrt{-\nabla ^2}}q,~~E_1=3D{1\over {{\sqrt{-\nabla =
^2}}}}(\epsilon_{ij}\partial _jp+\mu\partial _iq),$$
\begin{equation}
H=3D\int d^2x {1\over 2}(p^2+\partial _iq\partial _iq+\mu =
^2q^2),~~P^i=3D\int d^2x(p\partial ^iq).
\label{mcs}
\end{equation}
This set of relations is identical to those in \cite{djt} and hence
the results obtained by DJT will follow trivially.

The latter case, $\mu ^2=3D0$, refers to the Proca model, where =
$M_1^2=3DM_2^2=3Dm^2$, and we get,
$$
B=3D-{{\sqrt {-2\nabla ^2}}\over 2}(Q_1+Q_2);~~
E_i=3D-{1\over {\sqrt {-2\nabla ^2}}}[\epsilon _{ij}\partial =
_j(P_1+P_2)+m(\partial _iQ_1 -\partial _iQ_2)],
$$
$$
H=3D\int d^2x[{1\over 2}(P_1^2+\partial _iQ_1\partial =
_iQ_1+m_1^2Q_1^2)+{1\over 2}(P_2^2+\partial _iQ_2\partial =
_iQ_2+m_2^2Q_2^2)],$$
\begin{equation}
P^i=3D\int d^2x[P_1\partial ^iQ_1+P_2\partial ^iQ_2].
\label{fg}
\end{equation}
Following the prescription of DJT given in \cite{djt}, the boost =
generator $M^{i0}$ should be reinforced by the additional terms,
$$
m\epsilon_{ij}\int d^2x({{P_1\partial _jQ_1}\over{-\nabla ^2}}
-{{P_2\partial _jQ_2}\over{-\nabla ^2}}),$$
such that the electromagnetic fields transform correctly. This addition, =
however, generates a zero momentum anomaly in the boost algebra,
\begin{equation}
i[M^{i0},M^{j0}]=3D\epsilon^{ij}(M-\Delta),~~~\Delta=3D{{m^3}\over{4\pi =
}}
\{(\int Q_1)^2-(\int Q_2)^2\}+{m\over{4\pi }}\{(\int P_1)^2-(\int =
P_2)^2\},
\label{pr}
\end{equation}
where $M$ is the rotation generator=20
$$M=3D-\int d^2x(P_1\epsilon^{ij}x^i\partial =
_jQ_1+P_2\epsilon^{ij}x^i\partial _jQ_2)$$.
Making the mode expansions,
\begin{equation}
Q_1(x)(Q_2(x))=3D\int {{d^2k}\over{2\pi {\sqrt{2\omega =
(k)}}}}[e^{-ikx}a(k)(b(k))+e^{ikx}a^+(k)(b^+(k))],
\label{mo}
\end{equation}
and effecting the phase redefinitions,
\begin{equation}
a\rightarrow e^{i{m\over{\mid m\mid }}\theta }a,~~~b\rightarrow =
e^{-i{m\over{\mid m\mid }}\theta }b,
\label{ph}
\end{equation}
where $\theta =3Dtan^{-1}k_2/k_1 $, one recovers the full angular =
momentum as
\begin{equation}
M=3D\int d^2k(a^+(k){1\over i}{\partial \over {\partial \theta }}a(k)+
b^+(k){1\over i}{\partial \over {\partial \theta }}b(k))+{m\over {\mid =
m\mid }}\int d^2k(a^+(k)a(k)-b^+(k)b(k)),
\label{sp}
\end{equation}
where the second term is the spin.

Now comes the intriguing part, {\it i.e.} what happens when both $\mu $
and $m$ are nonzero. First of all, for simplicity, let us neglect
$O(\mu ^2)$ terms, which makes ${\cal H}_{St}$ a decoupled sum of "1"
and "2" variables. But even then, similar extensions in $M^{i0}$,
as done in the previous cases, will not have the desired effect since
the parameters present in ${\cal H}_{St}$, $M_1^2(M_2^2)\mid _{\mu =
^2=3D0}=3Dm^2\pm m\mu $ are different from the parameters appearing in =
the electric field, $(\mu \pm m )^2\mid _{\mu ^2=3D0}$. Obviously, if we =
keep the $O(\mu ^2)$ terms as well, the situation will worsen since =
${\cal H}_{St}$ is no longer decoupled.
This constitutes the main result of this Letter.

In an earlier work \cite{h}, it was argued that the anomaly in =
\cite{djt} appeared only because of the mapping of the system in terms =
of a scalar variable. However,  it has been demonstrated in \cite{djt} =
how to overcome this problem, leading to the correct spin value of the =
excitation in the process. As we have shown, this scheme is untenable in =
the MCSP model.

To conclude, We have shown that in the Maxwell-Chern-Simons-Proca model, =
where two mass scales, topological and non-topological or explicit, are =
present simultaneously, the electromagnetic field transforms anomalously =
under Poincare transformations. The conventional way \cite{djt} of =
redefining the phases of the creation and annihilation operators of the =
basic fields to remove the anomaly is inadequate in the present case.
A deeper understanding of this pathological behaviour is necessary.
However, in applications of condensed matter physics,
where Poincare or Lorentz invariance is generally not a big issue, these =
models can still play an important role.

\end{document}